\pdfoutput=1
\documentclass[svgnames,dvipsnames,letterpaper, 10 pt, conference]{ieeeconf}  

\IEEEoverridecommandlockouts                              

\usepackage{pgf}
\usepackage{url}
\usepackage{amsmath} 
\usepackage{amssymb}  
\usepackage{tikz}
\usetikzlibrary{shapes}
\usepackage[]{xcolor}
\usepackage{textcomp}

\newcommand{\ch}{\ensuremath{c}} 
\newcommand{\m}{\ensuremath{m}} 
\newcommand{\s}{\ensuremath{s}}
\newcommand{\x}{\ensuremath{x}} 
\renewcommand{\S}{\ensuremath{\bar{S}}} 
\newcommand{\D}{\ensuremath{\bar{D}}} 
\newcommand{\A}{\ensuremath{\bar{A}}} 

\title{\LARGE \bf
Minimizing subject-dependent calibration for BCI\\with Riemannian transfer learning
}

\author{Salim Khazem$^{1}$, Sylvain Chevallier$^{1}$, Quentin Barth\'elemy$^{2}$, Karim Haroun$^{1}$ and Camille No\^us$^{3}$
\thanks{$^{1}$Salim Khazem, Sylvain Chevallier and Karim Haroun are with LISV, Universit\'e de Versailles Saint-Quentin,
        University Paris-Saclay, Saclay, France
        {\tt\small sylvain.chevallier@uvsq.fr}}%
\thanks{$^{2}$Quentin Barth\'elemy is with Foxstream,
        Vaulx-en-Velin, France
        {\tt\small q.barthelemy@foxstream.fr}}%
\thanks{$^{3}$Camille No\^us is with Cogitamus,
        CNRS, France
        {\tt\small camille.nous@cogitamus.fr}}
}

\begin{document}

\maketitle
\thispagestyle{empty}
\pagestyle{empty}

\begin{abstract}

Calibration is still an important issue for user experience in Brain-Computer Interfaces (BCI). Common experimental designs often involve a lengthy training period that raises the cognitive fatigue,  before even starting to use the BCI. Reducing or suppressing this subject-dependent calibration is possible by relying on advanced  machine learning techniques, such as transfer learning. Building on Riemannian BCI, we present a simple and effective scheme to train a classifier on data recorded from different subjects, to reduce the calibration while preserving good performances. The main novelty of this paper is to propose a unique approach that could be applied on very different paradigms. To demonstrate the robustness of this approach, we conducted a meta-analysis on multiple datasets for three BCI paradigms: event-related potentials (P300), motor imagery and SSVEP. Relying on the MOABB open source framework to ensure the reproducibility of the experiments and the statistical analysis, the results clearly show that the proposed approach could be applied on any kind of BCI paradigm and in most of the cases to significantly improve the classifier reliability. We point out some key features to further improve transfer learning methods.
\end{abstract}

\section{Introduction}

Brain-computer interfaces allow to interact with a computer or a robotic system without relying on muscular capacity. 
As EEG measurements could be made with relatively inexpensive or easily available devices when compared to other kind brain imaging techniques, it is a common choice to conduct studies on many subjects.
While the BCI research is very active and diverse, the EEG-based BCI still suffers from important limitations~\cite{thompson_critiquing_2018}. 
Thus, they are mainly used for lab experiment and practical applications are scarce. 
Indeed, it is possible to enhance the preprocessing or the classification algorithms to better handle the individual variability~\cite{lotte_review_2018}.
Advanced signal processing and machine learning techniques are available but could not reduce the EEG-based BCI limitations completely.

An important leverage for improving the reliability of BCI is to improve the human-computer interfaces aspects~\cite{lotte_bcichallenge_2018}. 
The existing protocols are suboptimal and could be enhanced~\cite{jeunet_why_2016}.
To address these issues, it is possible to build better model of the user and to better characterize what the user is learning~\cite{lotte_bcichallenge_2018}.
One of the main sources of EEG variability is the user, due to his/her fatigue, mood change or attention variations. 

The BCI protocols generate an important cognitive fatigue, especially if the user is maintaining a high attentional focalization during the interactions.
To enforce the BCI reliability, it is common to include a calibration task for training preprocessing and machine learning algorithms.
This calibration could be long; it is a source of fatigue and distraction, even before the starting to use the BCI.
While setting up a subject independent calibration or a calibration-free system~\cite{fazli2009subject} is possible, the performances are not as good as subject-dependent calibration systems. 

To reduce -- and ideally to remove -- calibration it is possible to rely on transfer learning.
This is an important topic in machine learning and many approaches have been adapted and investigated in BCI~\cite{lotte_review_2018}.
As for classification algorithms, Riemannian approaches that rely on covariance matrices, have achieved robust and efficient discrimination~\cite{yger2017riemannian,congedo2017riemannian,Jayaram2019}.
This geometrical approach of the EEG feature representation allows to define metric spaces with chosen invariances.
One main interest of the so-called Riemannian BCI is that the same classification algorithms could be applied to virtually any type of BCI.
Indeed, transfer learning using covariance matrices has been investigated and achieves high performance~\cite{rodrigues2019transfer}, but nonetheless requires calibration data.

The proposed algorithm named MDWM, grounded in the manifold of covariance matrices, extends the works presented in~\cite{kalunga2018transfer} on SSVEP. 
Our contributions are twofold: MDWM works with small number of training data and it could be applied for different BCI paradigms.
In this paper, we generalize the MDWM transfer learning approach to different paradigms and we conduct a meta-analysis on P300 datasets, Motor Imagery datasets and SSVEP datasets.
After a short presentation of the method in Section~\ref{sec:methods}, this approach is evaluated on multiple datasets in Section~\ref{sec:results}.
Section~\ref{sec:conclusion} concludes this report.


\section{Methods}
\label{sec:methods}

\subsection{Riemannian BCI}
\label{sec:riemannianBCI}

Common preprocessing and classification algorithms for EEG-based BCI rely on estimation of covariance matrices.
These matrices are symmetric positive-definite and are thus restricted on a specific region of the space of real matrices.
Furthermore, these matrices have a characteristic geometry and, endowed with a proper metric, it is possible to benefit from invariance properties to make robust classification~\cite{congedo2017riemannian}.

The most simple yet effective Riemannian classifier for BCI is the minimum distance to mean (MDM).
This algorithm brings a significant improvement when compared to Euclidean metrics~\cite{CHE18} but does not achieve the highest classification performance amongst the Riemannian approaches~\cite{jayaram2018moabb}.
Most notably, MDM is robust to noise and works well with online implementation~\cite{kalunga_online_2016}.
This is thus a very good choice to serve as a basis for a transfer learning approach that should be used with minimal number of calibration samples.

From a set of $\left\{ \x_1, \ldots, \x_m \right\}$ EEG signals with $\ch$ channels extracted from $\m$ trials, we can estimate the $\left\{ \s_1, \ldots, \s_m \right\}$ associated covariance matrices, using $s_i = 1/\ch \ x_i^\intercal x_i$.
It is advised to use more advanced covariance estimators, such as described in~\cite{CHE18}.
In the space of covariance matrices, the shortest curve between two matrices $\s_i$ and $\s_j$ is known as the geodesic:
\begin{equation}
    \s_i \#_\lambda \s_j = \s_i^{\frac{1}{2}} \left( \s_i^{-\frac{1}{2}} \s_j \s_i^{-\frac{1}{2}}\right)^\lambda \s_i^{\frac{1}{2}} \ ,
    \label{eq:geodesic}
\end{equation}
with $\lambda \in [0, 1]$.
The length of this geodesic is the Riemannian distance. 

The first step of the MDM is to compute the average $\S_k$ covariance matrix for each of the $K$ classes. 
Then, it is possible to assign a newly seen covariance matrix $\s$ to the class with the minimum distance to mean, in the sense of the Riemannian distance. 

\subsection{Transfer Learning}
\label{sec:transfer}

The proposed transfer learning approach consists in using as little as possible calibration data from a user, relying on augmented models that make use of information available from other users.
As the proposed approach combine difference source EEG to estimate any target EEG signal, this approach is called Minimum Distance to Weighted Mean (MDWM)~\cite{kalunga2018transfer}.

This is accomplished by combining the estimated centers of class using the data available from the target subject $\S_k$ and the data available from the dataset of source subjects $\D_k$,
with $\A_k = \S_k \#_\lambda \D_k$, using Eq.~\eqref{eq:geodesic} 
The parameter $\lambda$ controls the trade-off between target and source data.
For $\lambda = 0$, only the few data $\S_k$ from the target subject are used (ie, no transfer scheme).
For $\lambda = 1$, the calibration of the target subject is completely discarded and we rely entirely on $\D_k$ data extracted from the source subjects (ie, calibration free scheme).

The proposed algorithm, MDWM, use only a few calibration data by combining them with existing data from the source subjects.
A weight can be added for each source subjects, defined for example as the similarity with the target subject \cite{kalunga2018transfer}.
In this study, uniform weights are chosen throughout the experiments.

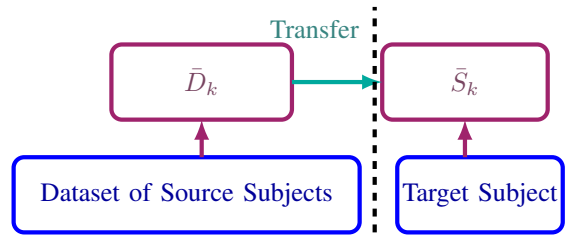
\begin{figure}[tb]
    \centering
    \begin{tikzpicture}
    \draw[ultra thick, color=blue, rounded corners] (0, 0) rectangle (4.6, 1) ;
    \draw[color=blue!60!black] (2.3, 0.5) node {Dataset of Source Subjects} ;
    \draw[ultra thick, color=blue, rounded corners] (5.1, 0) rectangle (7.3, 1) ;
    \draw[color=blue!60!black] (6.2, 0.5) node {Target Subject} ;
    
    \draw[>=latex, ->, ultra thick, color=RedViolet] (2.5, 1) --  (2.5, 1.5) ;
    \draw[>=latex, ->, ultra thick, color=RedViolet] (6, 1) --  (6, 1.5) ;
    
    \draw[ultra thick, color=RedViolet, rounded corners] (1.3, 1.5) rectangle (3.7, 2.5) ;
    \draw[color=RedViolet!60!black] (2.5, 2) node {$\D_k$} ;
    \draw[ultra thick, color=RedViolet, rounded corners] (4.9, 1.5) rectangle (7.1, 2.5) ;
    \draw[color=RedViolet!60!black] (6.0, 2.) node {$\S_k$} ;
    
    \draw[>=latex, ->, ultra thick, color=Emerald] (3.7, 2) --  (4.9, 2) ;
    \draw[color=Emerald!60!black] (4., 2.7) node {Transfer} ;
    
    \draw [dashed, ultra thick]  (4.8, 0) -- (4.8, 3.0);
    \end{tikzpicture}
    \caption{illustration of the cross-subject transfer learning: combining small calibration from the target subject with data extracted from source subjects.}
    \label{fig:pipelines}
\end{figure}

\subsection{MOABB}
\label{sec:moabb}
The choice of MOABB framework~\cite{jayaram2018moabb} is grounded in the need for reproducibility: this open-source project allows to download many EEG datasets and to easily evaluate any algorithm on those datasets\footnote{\url{https://github.com/NeuroTechX/moabb}}.
The objective of MOABB is to yield reliable machine learning framework and sound statistical analysis to compare classification algorithms.
Working on publicly available datasets, it is thus possible to reproduce results.

While MOABB is a powerful tool, it still lacks the support to do cross-subject transfer learning. It is a more global issue as no existing framework allows to fairly evaluate transfer learning algorithms.
This works is a first attempt to improve MOABB with transfer learning support and we will contribute to this open source project.
Our analysis focuses three kinds of paradigms, SSVEP, Motor Imagery (MI) and P300 Speller.
For the MI, the choice was based on two typical paradigms signal, left-hand vs right-hand imagery. 
The number of channels and subjects for each paradigm is described in Table~\ref{tab:datasets}.

\begin{table}[t]
    \centering
    \begin{tabular}{|c|l|c|c|c|}
        \hline
        Dataset name & Paradigm & \#Classes & \#Channels & \#Subjects \\
        \hline
        \hline
        SSVEP Exo \cite{kalunga_online_2016} & SSVEP & 4 & 8 & 12 \\
        Nakanishi 2015 \cite{nakanishi2015comparison} & SSVEP & 12 & 8 & 10 \\
        \hline
        \hline
        Zhou 2016 \cite{zhou_fully_2016}   & MI & 3 & 14 & 4 \\
        BNCI-004-2014 \cite{bnci0004-2014} & MI & 2 & 22 & 9 \\
        BNCI-001-2014 \cite{bnci0001-2014} & MI & 2 & 22 & 9 \\
        \hline \hline
        BNCI-008 2014 \cite{bnci0008-2014} & P300 & 2 & 8 & 8 \\
        BNCI-009 2014 \cite{bnci0009-2014} & P300 & 2 & 16 & 10 \\
        BNCI-003 2015 \cite{bnci0003-2015} & P300 & 2 & 8 & 10 \\
        \hline
    \end{tabular}
    \caption{List of datasets: SSVEP, Motor Imagery and P300 speller.}
    \label{tab:datasets}
\end{table}

We have benchmarked MDWM on 7 different datasets (2 SSVEP, 2 MI and 3 P300) and shows that it performs significantly better than reference machine learning pipelines (CCA, CSP+LDA and xDAWN+LDA).

\begin{figure}[tb!!]
  \centering
  \pgfimage[interpolate=true,width=\linewidth]{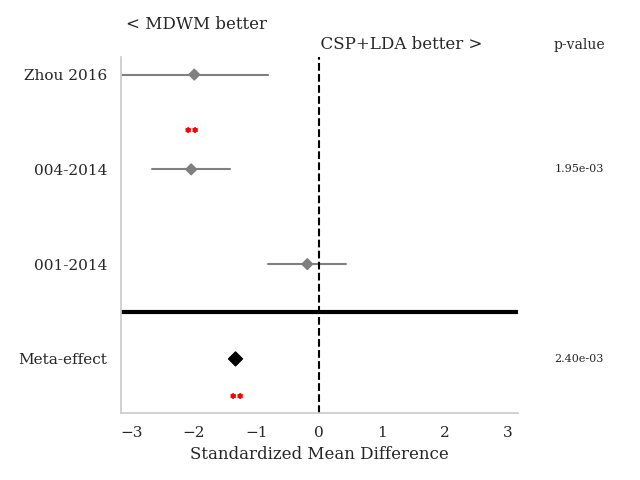} 
  \pgfimage[interpolate=true,width=\linewidth]{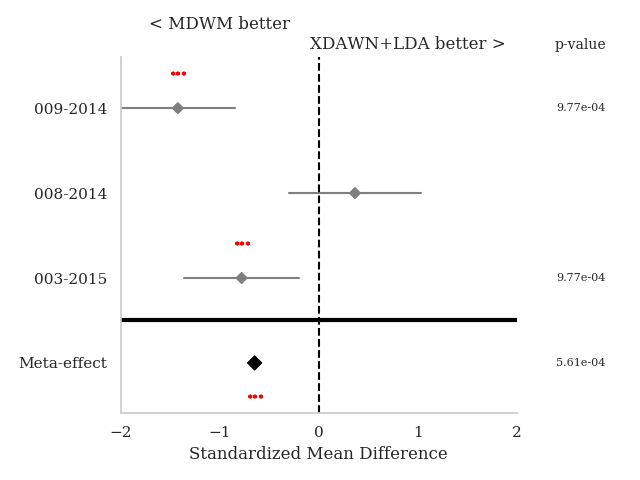}
  \pgfimage[interpolate=true,width=\linewidth]{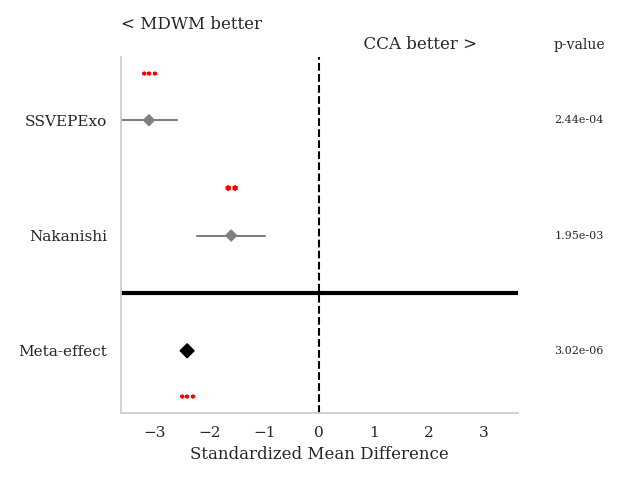}
  \caption{Meta-analysis on several datasets, for different BCI paradigms comparing MDWM against CSP+LDA on MI (top), with xDAWN+LDA on P300 (middle), and with CCA on SSVEP (bottom).}
  \label{fig:metaanalysis}
\end{figure}

\subsection{Analysis on transfer learning}

Our objective being to minimize the calibration step of the target subject by using data from source subjects. 
To evaluate the influence of the number of samples used from the target subject, we use different numbers of training samples from the target subject $n$ $\in$ $\left\{5, 30, 55\right\}$.
To evaluate the influence of the trade-off between source and target subjects, we report MDWM performances for different values of $\lambda$ $\in$ $\left\{0, 0.1, 0.3, 0.7\right\}$. 

For each dataset, the evaluation is conducted as follows:
the target subject data is separated from the other source subjects in a given dataset (following a leave-one-subject-out scheme),
the trials from the target subject are separated into $n$ trials for training (considered as calibration data) and the rest of the trials are used for testing.
The model is trained on all the data available from the source subjects and the $n$ trials from the target subject.
The model is evaluated on the rest of the trials from the target subject.
This process is repeated 10 times, with different trials kept as the $n$ training samples

\subsection{Meta-analysis on classification accuracy}
In order to understand if the proposed method (MDWM) is consistently better than the state-of-the-art method (\emph{i.e.}, CSP+LDA for MI, xDAWN+LDA for P300 and CCA for SSVEP) across datasets and paradigms, we use a meta-analysis.
It formulates a null hypothesis $H_0$, which we are trying to reject, that there is no difference between the two methods.
Then, we want to answer the following question: under $H_0$, what is the probability to obtain the difference in the means that we observed in the experimental data? 

To obtain results with comparable metrics across paradigms, all evaluations are performed with the same classification metric: the balanced accuracy, more appropriate when classes are imbalanced like in P300.
Meta-analysis gives two values: a standardized mean difference (SMD) and a p-value $p$.
SMD is a normalized measure to assess the difference between the means of the classification metrics obtained by the two methods. It allows to understand the classification delta between both methods.
The p-value is the probability of observing such difference considering the null hypothesis $H_0$ as true. It allows to reject $H_0$ when it is very low. 
Statistical significance is $\star$ for $p < 0.05$, $\star\star$ for $p < 0.01$ and $\star\star\star$ for $p < 0.001$, using Stouffer's method to combine $p$-values obtained with Wilcoxon signed-rank test on each separate dataset.



\section{Results and Discussions}
\label{sec:results}

In these experiments, we have used MDWM with $\lambda = 0.7$ and $n$ that is two times the number of classes. 
Results of meta-analysis are shown in Fig.~\ref{fig:metaanalysis}.
The top plot shows that MDWM outperforms CSP+LDA for MI in all three datasets, with a sizable SMD and a good statistical significance.
On middle plot, for the case of P300 speller paradigm, we can notice that MDWM outperforms xDAWN+LDA on two datasets over 3, with a noticeable SMD and a high statistical significance. 
On the bottom plot, MDWM outperforms CCA for both SSVEP datasets, with a high SMD and a high statistical significance.
These results are a clear validation of the proposed transfer learning method.

\begin{figure}[t]
  \centering
  \pgfimage[interpolate=true,width=\linewidth]{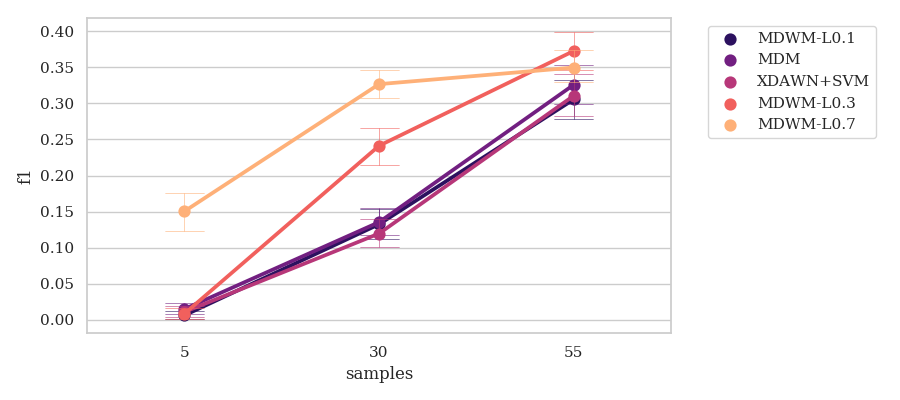}
  \caption{Classification accuracy depending on the number of samples available for P300 datasets.}
  \label{fig:classif}
\end{figure}

The Fig.~\ref{fig:classif} shows that by increasing the number of samples, the accuracy increases.
We notice that by having a $\lambda$ equal to 0.7, which correspond to rely mostly on data from source subjects and only moderately on target subject, we obtain fairly good results. 

As a discussion from these transfer learning results, we could report that not all subjects have an equivalent contribution for reducing the calibration.
The "transfer-ability" of a subject is an open question and there are interesting investigation on this matter in~\cite{rodrigues2019transfer}.
What is interesting here is that the MDWM is sufficiently robust to aggregate results from all subjects without distinction.
Indeed, a proper selection of source subjects to train the MDWM could really improve the performances.

We also investigate if the calibration could be completely bypassed, by make use of the resting state EEG.
It is common to start the EEG recording by asking the subject to close his/her eyes or to relax a little bit with eyes closed or opened.
This resting EEG could be used to characterize the similarity between two subjects and thus to select best matching subject to train MDWM.
We have investigated if resting EEG convey as much information as task related data for selecting similar subjects and it is the case.
This subject will be the focus of another study.

The performances of MDWM could be improved by carefully selecting the $\lambda$ value.
A possible choice is to set $\lambda$ as a function of the $n$, the number of calibration trial.
Thus, the more data is available from the subject, the lower could be $\lambda$.
We plan to investigate different function to set up automatically $\lambda$.

As we have pointed out, the reproducibility of results is a crucial issue for signal processing and machine learning approaches in BCI.
MOABB, or other open source frameworks, are important to ensure that a fair evaluation of existing algorithms is available.
While no framework proposes to benchmark transfer learning approaches yet, we will contribute improving MOABB on this point.


\section{Conclusion}
\label{sec:conclusion}

In this contribution, we have described a transfer learning approach called MDWM and we have demonstrated that it could be applied on three different BCI paradigms, namely motor imagery, P300 and SSVEP.
The MDWM uses a combination of a few calibration data from a target user and data available from other subjects.
We have benchmarked MDWM on 7 different datasets and shows that it performs significantly better than reference machine learning pipelines (CSP, xDAWN and CCA).
This is, to the best authors' knowledge, the first attempt to systematically evaluate one classification algorithm on three different BCI paradigms.
It should be noted that we did not optimize the $\lambda$ parameter to obtain these results. 
These results are promising and are a first step towards an enhancement of a framework for systematic and fair evaluation of transfer learning.





\section*{Environmental impact}

The approach taken in this submission tried to reduce the number of experiments made on servers, in order to reduce its environmental impact.
The authors communicated through Slack, git and overleaf.
Based on Shift Project and ADEME reports, we estimate that the impact of publication is roughly equivalent to 73 gCO$_2$.

\section*{Acknowledgment}

QB and SC would like to thank Emmanuel Kalunga for the discussions about this project.


\bibliographystyle{unsrt}
\bibliography{bibliobci.bib}

\end{document}